%% file: quasi-online.tex
\def\version{2}

\if\version0
\documentclass[12pt]{article}
\newcommand{\blind}{0}

\def\spacingset#1{\renewcommand{\baselinestretch}%
  {#1}\small\normalsize} \spacingset{1}
\fi

\if\version2
\documentclass[a4paper,11pt,reqno]{article}
\fi

\usepackage{mathtools}
\mathtoolsset{showonlyrefs}
\usepackage{bm}
\usepackage{siunitx}
\usepackage{booktabs}
\usepackage{microtype}
\usepackage{amsthm,amsmath,amsfonts,amssymb}
\usepackage{graphicx} 
\usepackage{xcolor}
\usepackage{float}
\usepackage{doi}
\usepackage[round]{natbib}
\usepackage{algorithm}
\usepackage[noend]{algpseudocode}
\algrenewcommand\algorithmicrequire{\textbf{Input:}}
\algrenewcommand\algorithmicensure{\textbf{Output:}}
\usepackage{enumerate}
\usepackage{fullpage}
\usepackage{xspace}
\usepackage{stmaryrd}
\usepackage{placeins} 
\usepackage[normalem]{ulem}
\usepackage{subcaption}
\usepackage{paralist}

\usepackage[top=2.5cm, bottom=2.5cm, left=2.5cm, right=2.5cm]{geometry}	%Margins

\if\version2
\mathtoolsset{showonlyrefs} % number only equations that are referenced
\usepackage{hyperref}
\fi

\hypersetup{
  colorlinks   = true, %Colours links instead of ugly boxes
  urlcolor     = {green!50!black}, %Colour for external hyperlinks
  linkcolor    = {black}, %Colour of internal links
  citecolor   = {green!50!black} %Colour of citations
}

\usepackage{ifthen}%
\usepackage{mfirstuc}
\usepackage{xkeyval}%
\usepackage{xfor}%
\usepackage{amsgen}%
\usepackage{etoolbox}%
\usepackage{datatool-base}%

\usepackage[%
  xindy,%
  style=long,%
  toc=true,%
  nonumberlist,%
  acronym,%
  acronymlists={abbreviations},%
  hyperfirst=true% link first use of abbreviation
  ]{glossaries}%
\input{abbreviations.tex}%

\usepackage[textwidth=1.8cm, textsize=scriptsize]{todonotes}
\usepackage{setspace} % only used for comments

\newcommand{\naturals}{\mathbb{N}}
\newcommand{\integers}{\mathbb{Z}}
\newcommand{\reals}[1][]{\mathbb{R}^{#1}\xspace}
\newcommand{\diff}{\mathrm{d}}
\newcommand{\intDiff}{\,\diff}

\newcommand{\dcount}[1]{\ensuremath{\llbracket #1 \rrbracket}\xspace}
\newcommand{\dinterval}[2]{\ensuremath{\llbracket #1,#2\rrbracket}\xspace}

\newcommand{\dnorm}{\mathcal{N}}
\newcommand{\ccdot}{\,\cdot\,}

\newcommand{\ess}[1]{\ensuremath{\mathop{\textsc{ess}}(#1)}}
\newcommand{\resample}[1]{\ensuremath{\mathop{\textsc{resample}}(#1)}}
\newcommand{\nParticles}{N}
\newcommand{\particleSystem}[1]{H_{#1}}
\newcommand{\particleSystemAlt}[1]{\tilde{H}_{#1}}

\newcommand{\particleSystemDefinition}[1]{\smash{(X_{#1}^{1:\nParticles}, W_{#1}^{1:\nParticles}, A_{#1-1}^{1:\nParticles}, Z_{#1}^\nParticles)}}

\newcommand{\twistingFunction}[1]{\psi_{#1}}

\newcommand{\psiapf}[1]{\mathop{\text{$\psi$-\textsc{apf}}}(#1, \twistingFunction{#1}, \particleSystem{#1 - 1})}
\newcommand{\psiapfAlt}[1]{\mathop{\text{$\psi$-\textsc{apf}}}(#1, \twistingFunction{#1}, \particleSystemAlt{#1 - 1})}

\newcommand{\learnpsi}[1]{\mathop{\text{\textsc{learn}-$\psi$}}(#1, \twistingFunction{#1 + 1}, \particleSystem{#1})}
\newcommand{\learnpsiAlt}[1]{\mathop{\text{\textsc{learn}-$\psi$}}(#1, \twistingFunction{#1 + 1}, \particleSystemAlt{#1})}
\newcommand{\transitionIntegral}[2]{f_{#1}(\twistingFunction{#1})(#2)}

\newcommand{\actualacknowledgements}{We are grateful to the Scientific Computing Research Technology Platform (SCRTP) at the University of Warwick for the provision of computational resources. LX acknowledges support from China Scholarship Council via grant number 202208060348. AMJ acknowledges support from the United Kingdom Engineering and Physical Sciences Research Council (EPSRC; grants EP/R034710/1 and EP/T004134/1) and from United Kingdom Research and Innovation (UKRI) via grant number EP/Y014650/1, as part of the ERC Synergy project OCEAN.\\\noindent\textbf{Data availability:} All data and code used in this paper can be obtained from \url{https://github.com/Sempreteamo/orc.smc/tree/main}.}
\newcommand\actualtitle{Online Rolling Controlled Sequential Monte Carlo}

\begin{document}
\if\version0
\if0\blind
{
  \title{\bf \actualtitle}
  \author{Liwen Xue$\vphantom{e}^1$, Axel Finke$\vphantom{e}^2$ and Adam M. Johansen$\vphantom{n}^1$\thanks{
    \actualacknowledgements }\hspace{.2cm}\\
    $\vphantom{e}^1$ --- Department of Statistics, University of Warwick, Coventry, UK\\
    $\vphantom{e}^2$ --- School of Mathematics, Statistics and Physics, Newcastle University, UK}
  \maketitle
} \fi

\if1\blind
{
  \bigskip
  \bigskip
  \bigskip
  \begin{center}
    {\LARGE\bf Title}
\end{center}
  \medskip
} \fi

\bigskip
\fi
\if\version2
\title{\actualtitle}
\author{Liwen Xue, Axel Finke and Adam M. Johansen}
\maketitle
\fi

\begin{abstract}
  \noindent{}We introduce methodology for real-time inference in general-state-space hidden Markov models. Specifically, we extend recent advances in controlled sequential Monte Carlo (CSMC) methods---originally proposed for offline smoothing---to the online setting via a rolling window mechanism. Our novel online rolling controlled sequential Monte Carlo (ORCSMC) algorithm employs two particle systems to simultaneously estimate twisting functions and perform filtering, ensuring real-time adaptivity to new observations while maintaining bounded computational cost. Numerical results on linear-Gaussian, stochastic volatility, and neuroscience models demonstrate improved estimation accuracy and robustness in higher dimensions, compared to standard particle filtering approaches. The method offers a statistically efficient and practical solution for sequential and real-time inference in complex latent variable models.
\end{abstract}

\noindent%
\if\version0
\vfill

\newpage
\spacingset{1.75} % DON'T change the spacing!
\fi

\glsunset{PSIAPF}

\section{Introduction}

This paper is concerned with inference for general-state-space \emph{\glspl{HMM}}, a.k.a.\ \emph{state-space models}. These models are used in diverse applications ranging from control systems to financial analysis, navigation, and biological monitoring. In such applications, efficient and accurate estimation of the hidden states (given noisy and partial observations) is crucial. Three inferential tasks are of particular interest:
\begin{compactenum}
    \item \textit{Filtering:} characterizing the conditional law of the most recent latent state given all the observations received thus far.
    \item \textit{Smoothing:} characterizing the conditional law of all latent states until the current time point given all the observations received thus far.
    \item \textit{Marginal-likelihood evaluation:} evaluating the density of all available observations with the states marginalised out. This is typically needed for performing likelihood-based estimation of unknown model parameters.
\end{compactenum}
Such inference is typically \emph{sequential} in the sense that beliefs about the latent states are to be updated whenever new observations become available---often at short intervals; and may even need to be \emph{online} in the sense that both the space and time costs of incorporating one new observation are bounded uniformly in time.
 
Obtaining the filtering and smoothing distributions and marginal likelihoods involves computing integrals which are typically intractable, except in a limited number of scenarios such as linear Gaussian models---which give rise to the celebrated Kalman filter and related recursions \citep{smc:hmm:Kal60}---and a few others \citep{smc:hmm:Vid99}. Consequently, Monte Carlo methods are the state of the art for general inference in these methods.

\emph{\Glspl{PF}} are a class of \emph{\gls{SMC}} methods adapted to conducting inference in general-state-space \glspl{HMM} (with a particular focus on online solutions of the filtering problem). They first came to prominence in the form of the \emph{\gls{BPF}} introduced in \citet{gordon1993novel}. Since then, there have been numerous innovations in the area. Those most critical to the current work are summarised in the next section; see, e.g., \citet{amj26:DJ11} for a survey of approaches to filtering and smoothing, \citet{smc:tutorial:KDSMC15} for an overview of parameter estimation techniques or \citet{smc:tutorial:CP20} for a book-length introduction to \gls{SMC} methods including their applications to general-state-space \glspl{HMM}. 

\Glspl{PF} perform poorly in challenging scenarios such as higher-dimensional latent states. To alleviate this problem, \citet{amj26:GJL17, smc:methodology:HBDD20} proposed \emph{\gls{CSMC}} methods (a.k.a., \emph{iterated auxiliary particle filters}) which leverage insights from optimal control theory to improve the efficiency of the \gls{PF}. Unfortunately, as \gls{CSMC} must repeatedly browse through the entire observation sequence, it is limited to \emph{offline} inference (i.e., settings in which the full data record is available before inference is conducted). 

In this work, we propose a novel method, \emph{\gls{ORCSMC}}, which aims to extend the benefits of \gls{CSMC} to online inference. \gls{ORCSMC} propagates a particle system using a finite-horizon modifications of the \gls{CSMC} approach. At each time point, aditional simulations temporarily extend this system to the time of the current observation in order to compute efficient approximations of filtering and smoothing distributions and marginal likelihoods in real time. \gls{ORCSMC} is also applicable in contexts in which storing on processing the entire data set (as required for \gls{CSMC}) is impractical.

Section~\ref{sec:background} reviews existing \gls{SMC} approaches for filtering and smoothing, as well as \gls{CSMC}; Section~\ref{sec:cops} presents the novel \gls{ORCSMC} method. Section~\ref{sec:experiments} demonstrates the performance of our proposed approach on different models;  Section~\ref{sec:conclusion} concludes.

Throughout this work, we use the shorthand $\dinterval{m}{n} \coloneqq \{k \in \integers\mid m \leq k \leq n \}$, for $m, n \in \mathbb{Z} \cup \{\infty\}$ with $m \leq n$. If $n \geq 1$, we also write $\dcount{n} \coloneqq \dinterval{1}{n}$. If $m$ and $n$ are finite, we further write $z_{m:n} \coloneqq (z_m, z_{m + 1}, \dotsc, z_n)$ and $z^{m:n} \coloneqq (z^m, z^{m+1}, \dotsc, z^n)$.

\section{Background}\label{sec:background}
This section reviews \gls{SMC} methods for inference in general-state-space \glspl{HMM}, with a focus on controlled methods and techniques designed for approximating smoothing distributions.

\subsection{General-state-space HMMs}

For definiteness, consider an $\reals[d]$-valued discrete-time Markov process $\{X_t\}_{t \geq 1}$ such that 
\begin{equation}
  X_1 \sim \mu \quad \text{and} \quad X_t| (
  X_{t-1} = x_{t-1}) \sim f_t(\ccdot|x_{t-1}),
\label{eq:evol}
\end{equation}
where $t > 1$, ``$\sim$'' means distributed according to, $\mu(x_1)$ is a probability density function and $f_t(x_t|x_{t-1})$ denotes the probability density associated with moving from $X_{t-1} = x_{t-1}$ at time~$t-1$ to $X_t = x_t$ at time~$t$. To simplify the notation, any additional `static' model parameters parametrising these densities are omitted from the notation. 

We are interested in performing inference about $\{X_t\}_{t \geq 1}$ but only have access to $\reals[d']$-valued observations $\{Y_t\}_{t \geq 1}$, assumed to be conditionally independent given $\{X_t\}_{t \geq 1}$ with densities
\begin{equation}
    Y_t | (X_t = x_t) \sim g_t( \ccdot | x_t).  \label{eq:obs}
\end{equation}
The main inference problems for this model class is characterizing the filtering and smoothing densities and marginal likelihoods:
\begin{align}
  p(x_t|y_{1:t}) &\propto \int \biggl[ \mu(x_1) g_1(y_1|x_1) \prod_{s=2}^t f_s(x_s|x_{s-1}) g_s(y_s|x_s)\biggr] \intDiff x_{1:t-1},\\
  p(x_{1:t}|y_{1:t}) & \propto \mu(x_1) g_1(y_1|x_1) \prod_{s=2}^t f_s(x_s|x_{s-1}) g_s(y_s|x_s), \label{eq:joint_smoothing_distribution}\\
  p(y_{1:t}) & = \int \biggl[\mu(x_1) g_1(y_1|x_1) \prod_{s=2}^t f_s(x_s|x_{s-1}) g_s(y_s|x_s)\biggr] \intDiff x_{1:t}.
\end{align}
If online inference is required, then the most we can generally hope to achieve is characterizing fixed-dimensional marginals of the smoothing distributions from \eqref{eq:joint_smoothing_distribution}.

\subsection{Particle Filtering}
\glsreset{BPF}
The \emph{\gls{BPF}} \citep{gordon1993novel} proceeds by sampling a collection of Monte Carlo samples (`particles') from the initial distribution $\mu$, weighting them in accordance with the likelihood associated with the first observation, $y_1$, and then---at time $t$---iteratively sampling from the resulting weighted empirical distribution, extending the sampled paths according to the transition density $f_t$ and weighting them with the likelihood of $y_t$.

Sampling from the weighted empirical distribution can be decomposed as a selection step, known as \emph{resampling}, in which particles with larger weights tend to produce more offspring followed by a mutation step in which particles move according to the underlying dynamic model. Seen in this way, the algorithm associates particle values at each time with its predecessor and introduces a genealogical relationship allowing us to think about the ancestor of a given particle at any number of generations earlier. Tracing ancestries back in this way is central to the most basic uses of \glspl{PF} in order to address smoothing problems.

The \gls{BPF} is effective in many settings, but has a number of limitations which more recent work has sought to overcome. 
More general sequential importance resampling \glspl{PF} \citep[see, e.g.,][]{smc:tutorial:DGA00} follow from recognising that the \gls{BPF} is essentially carrying out a sequence of importance sampling operations and hence that one can adjust the proposal distribution to better take into account the observation at the current time, with a proposal distribution of $p(x_t|x_{t-1},y_t)$ providing the smallest conditional variance for the importance weights at step $t$, the appropriate weights being proportional to $p(y_t|x_{t-1})$ in this case.

The \emph{\gls{APF}} approach is a further enhancement of standard \glspl{PF}. Originally introduced using auxiliary variables, it improves the effectiveness of state estimation by incorporating a look-ahead mechanism into the resampling procedure which predicts the likelihood that particles will be in regions of high probability at the next time step \citep{pitt1999filtering,smc:methodology:CCF99}. This algorithm can also be understood as a variant of sequential importance resampling \citep{amj26:JD08} which targets a slightly different sequence of distributions and corrects for the discrepancy between those and the filtering distributions with an additional importance sampling correction.

While simple \glspl{PF} are often able to provide adequate approximations of filtering distributions, more specialised techniques are generally required in smoothing contexts. \glspl{PF} suffer from limitations such as particle impoverishment in which repeated resampling operations lead to only a small number of distinct state values at early times having descendants at later times and weight degeneracy in which the importance weights have very high variance with only a small number differing substantially from zero; these and related phenomena are particularly problematic when the goal is to approximate smoothing distributions. 

\subsection{Particle Smoothing: Fixed-lag and Other Schemes} 

In general, one seeks good sample approximations of the joint distribution $p(x_{1:T}|y_{1:T})$ and its marginals ${p(x_t|y_{1:T})}$, for $t \in \dcount{T}$, and in the online context one similarly seeks approximations of $p(x_{1:t}|y_{1:t})$ and its marginals $p(x_s|y_{1:t})$, for $s \in \dcount{t}$, which can be updated as the $t$th observation arrives, for $t = 1, \dotsc, T$. \emph{Online filtering} as described in the previous section focuses on real-time estimation, while \emph{offline smoothing} involves utilizing the complete set of observations to refine state estimates. \emph{Online smoothing}, as distinct from offline approaches, is concerned with making estimates as new data arrives while retaining a focus on both current and prior states. 

Online smoothing in which we obtain good approximations of the smoothing distributions iteratively as we obtain observations at a computational cost which does not grow unboundedly with time is a challenging problem and perhaps the main focus of this work. Often, one may wish to approximate the marginals ${p(x_s|y_{1:t})}$ for $s \in \dcount{t}$. However, when $t$ is large, the simple approach (sometimes termed the ``smoothing mode of the particle filter'') in which one just traces back the histories of the surviving particles at time $t$ to obtain this approximation generally fails also a result of the degeneracy of this sample approximation. Although \cite{smc:tutorial:Kit14} demonstrates that one can obtain good estimation if one uses very large numbers of particles in this context, approaches which require less substantial computational resources are needed.

The fixed-lag \gls{SMC} approach from \citet{kitagawa2001monte} is a practical method for smoothing in dynamic systems, which aims to balance computational efficiency with estimation accuracy. It was introduced to tackle the inefficiency of standard smoothing in scenarios where observations continue to arrive over time. For certain \glspl{HMM} with `forgetting' properties, for a sufficiently large lag $L > 1$, observations beyond $s + L$ provide minimal additional information about the trajectory $X_{1:s}$. We may then approximate $p(x_{1:s} | y_{1:t})$ by $p(x_{1:s} | y_{1: \min\{s + L, t\}})$. However, the choice of $L$ involves a bias--variance trade-off: If $L$ is too small, the model's memory has not dissipated, causing bias. Conversely, an excessively large $L$ leads to path degeneracy in the \gls{PF} and high variance.  \cite{olsson2008sequential} proposes an optimal lag choice of $\lceil c \log (s)\rceil$ under specific mixing conditions, noting that in practice, $L$ values between 20 and 50 often suffice. This method approximates only marginal distributions $p(x_s|y_{1:t})$, not the joint distribution $p(x_{1:t}, y_{1:t})$, and does not converge to the true distribution as $N \to \infty$, but this fixed-lag approach lays the foundation for more sophisticated methods.

Motivated by similar considerations, \citet{smc:methodology:LCL13} incorporate information from subsequent observations into proposal distributions in a principled manner. This can substantially improve performance but limits use in online settings as estimation is delayed until a number of additional observations are received; it also requires careful tuning of approximations of certain conditional distributions associated with the algorithm which is challenging for most realistic models. \citet{doucet2006efficient} address the first of these issues by developing a related algorithm---amenable to online application---in which a block of recent states are refreshed as each new observation becomes available. But again good approximations of some complex conditional distributions must be obtained by hand in order to implement this algorithm.

Finally, considerable effort has been dedicated to developing Monte Carlo schemes for addressing the smoothing problem. However, these methods tend to either involve a forward-backward structure making them unsuitable for online applications (such as the methods of \citet{smc:tutorial:BDM10}), or provide approximations of only single time marginals of the smoothing distribution \citep[e.g.,][]{fearnhead2010sequential}. The PaRIS algorithm of \citet{smc:methodology:OW17} and related methods are one promising recent approach but are restricted to approximating additive functionals.

\subsection{$\psi$-APF}
\glsreset{PSIAPF}

Before describing \glsdesc{CSMC}, the algorithm we seek to adapt to an online context, we introduce the algorithm that sits at its heart which was introduced as the \emph{\gls{PSIAPF}} by \cite{amj26:GJL17}. Underlying this algorithm is the construction of a class of \emph{twisted} \glspl{HMM}, parameterized by a sequence $\psi \coloneqq \{\twistingFunction{t}\}_{t\geq1}$ of continuous and positive functions $\twistingFunction{t} \colon \reals[d] \to (0, \infty)$. In describing these, it is convenient to introduce the integral notation $\mu(\twistingFunction{1}) \coloneqq \int \mu(x_1) \twistingFunction{1}(x_1) \intDiff x_1$ and $\transitionIntegral{t}{x_{t-1}} \coloneqq \int f_t( x_t|x_{t-1}) \psi_t(x_t) \intDiff x_t$, for $t > 1$. These twisted models adjust the dynamics and observation likelihoods of the original \gls{HMM} to account for future information via a form of importance weighting.

To simplify notation in the boundary case $t = 1$, we define $f_1(x_1|x_0) \coloneqq \mu(x_1)$ with the convention that any quantity with time subscript $0$ should be ignored. Thus, we can write $\mu(\twistingFunction{1}) = f_1(\twistingFunction{1})(x_0) = f_1(\twistingFunction{1})$ which makes the presentation of algorithms simpler. 

Specifically, for $t > 1$, the twisted models are:
\begin{align}
    \mu^{\psi}(x_1)
    & = f_1^\psi(x_1) \coloneqq \frac{\mu(x_1) \twistingFunction{1}(x_1)}{\mu(\twistingFunction{1})}, & 
    f_t^\psi(x_t|x_{t-1}) 
    & \coloneqq \frac{f_t(x_t|x_{t-1}) \twistingFunction{t}(x_t)}{\transitionIntegral{t}{x_{t-1}}},\\
    g_1^{\psi}(x_1)
    &\coloneqq \frac{g_1(y_1|x_1) \mu(\twistingFunction{1})}{\twistingFunction{1}(x_1)}  \transitionIntegral{2}{x_1}, &
    g_t^{\psi}(x_t)
    & \coloneqq \frac{g_t(y_t|x_t)}{\twistingFunction{t}(x_t)} \transitionIntegral{t+1}{x_t}. \label{eq:twisted_potential_function}
\end{align}
Given the sequence $\psi$, one simply applies a \gls{BPF} to the twisted model defined by $\{(f_{t}^\psi, g_{t}^\psi)\}_{t \geq 1}$. The resulting algorithm, the \gls{PSIAPF}, is obtained by applying Algorithm \ref{alg:psiapf} recursively for $t = 1, 2, \dotsc$

\begin{algorithm}[htpb]
\caption{$\psiapf{t}$}
\label{alg:psiapf}
\begin{algorithmic}[1]
 \Require Time index $t \in \naturals$.
 \Require Twisting function $\twistingFunction{t}$.
 \Require Particle system $\particleSystem{t-1}$ as in \eqref{eq:particle_system}, if $t > 1$; if $t = 1$, then $\particleSystem{0}$ is as in \eqref{eq:initial_particle_system}.
  \State\label{alg:psiapf:first_weight_update:1}Set $v^n \coloneqq W_{t-1}^n \transitionIntegral{t}{X_{t-1}^n}$, for $n \in \dcount{N}$. 
  \State\label{alg:psiapf:first_weight_update:2}Set $Z_t^\nParticles \coloneqq Z_{t-1}^\nParticles \sum_{n=1}^N v^n$ and self-normalise: $V^n \coloneqq v^n / \sum_{m=1}^N v^m$, for $n \in \dcount{N}$.
  \If {$\ess{V^{1:\nParticles}} < \kappa N$} \label{alg:psiapf:resample:1} \Comment{$\smash{\ess{V^{1:\nParticles}} \coloneqq 1 / \sum_{n=1}^N (V^n)^2}$}
    \State\label{alg:psiapf:resample:2}$A_{t-1}^{1:\nParticles} \leftarrow \resample{V^{1:\nParticles}}$, 
    \State\label{alg:psiapf:resample:3}$V^n \leftarrow 1/N$, for $n \in \dcount{N}$;
  \Else
      \State\label{alg:psiapf:resample:4} $A_{t-1}^{1:\nParticles} \leftarrow (1, \dotsc, N)$.
  \EndIf
  \State Sample $\smash{X_t^n \sim f_t^\psi(\ccdot | X_{t-1}^{A_{t-1}^n})}$, for $n \in \dcount{N}$.
  \State\label{alg:psiapf:second_weight_update:1}Set $w_t^n \coloneqq V^n \smash{g_t(y_t|X_t^n) / \psi_t(X_t^n)}$, for $n \in \dcount{N}$. 
  \State\label{alg:psiapf:second_weight_update:2}Set $Z_t^\nParticles \leftarrow Z_t^\nParticles \sum_{n=1}^N w_t^n$ and self-normalise: $W_t^n \coloneqq w_t^n / \sum_{m=1}^N w_t^m$, for $n \in \dcount{N}$.
 \Ensure Particle system $\particleSystem{t} = \particleSystemDefinition{t}$.
\end{algorithmic}
\end{algorithm}

Algorithm~\ref{alg:psiapf} can be viewed as (the $t$th step of) a form of \gls{APF} \citep{pitt1999filtering}. It also generalises the \gls{BPF} (with $\twistingFunction{t} \equiv 1$, for $t \geq 1$) and the fully-adapted \gls{APF} (with $\twistingFunction{t} \coloneqq g_t(y_t|\ccdot)$, for $t \geq 1$). Some comments about Algorithm~\ref{alg:psiapf} are in order. 

Firstly, to keep the notation concise throughout, we have defined the \emph{particle system} generated by the \gls{PSIAPF} at time $t$ as
\begin{align}
    \particleSystem{t} = \particleSystemDefinition{t}, \label{eq:particle_system}
\end{align}
where $X_t^n \in \reals[d]$ is the $n$th particle, $W_t^n \in [0,1]$ is its self-normalised weight, $A_{t-1}^n \in \dcount{\nParticles}$ the index of its ancestor particle which is generated by the resampling procedure (in other words, $\smash{X_{t-1}^{A_{t-1}^n}}$ is $n$th resampled particle from time $t-1$), and $Z_t^\nParticles$ is used to construct an estimate of the normalising constant of the twisted \gls{HMM}; specifically, if the \gls{PSIAPF} is iterated up to any finite final time $T$, in the sense that the twisting functions include information from observations only over $t \in \dcount{T}$, $Z_T^\nParticles$ is an unbiased estimate of $p(y_{1:T})$ \citep[Proposition 1]{amj26:GJL17}. Actually, as written, $Z_t^\nParticles$ provides an unbiased estimate of $p(y_{1:t})$ for \emph{any} $t$, but the variance of this estimate will typically be large if the twisting functions have incorporated information from subsequent observations, e.g., from $y_{t+1}$. For the particle system at time $t = 0$, we use the convention 
\begin{gather}
 W_0^1 = \dotsb = W_0^\nParticles \equiv 1/\nParticles, \quad \text{and} \quad Z_0^\nParticles \coloneqq 1.\label{eq:initial_particle_system}
\end{gather}
The values of $\smash{X_0^{1:\nParticles}}$ and $\smash{A_0^{1:\nParticles}}$ can be arbitrary as they are not used by the algorithm.

Secondly, Lines~\ref{alg:psiapf:first_weight_update:1}--\ref{alg:psiapf:first_weight_update:2} update the time-($t-1$) particle weights (and normalising-constant estimate) to account for the second term, $\transitionIntegral{t}{\ccdot}$, in \eqref{eq:twisted_potential_function} (recall that this term is simply the scalar $\mu(\twistingFunction{1})$ if $t = 1$). This differs from the presentation of the \gls{PSIAPF} in \citet{amj26:GJL17, smc:methodology:HBDD20} which views this update as happening at the end of Step~$(t-1)$ rather than at the beginning of Step~$t$ but will be convenient for the online methodology developed in the next section. Specifically, this formulation allows us to carry out the ($t-1$)th step of the algorithm even if $\twistingFunction{t}$ is not yet known. However, we stress that the difference is purely presentational, i.e., combining Lines~\ref{alg:psiapf:first_weight_update:1}--\ref{alg:psiapf:first_weight_update:2} from Step~$t$ with Lines~\ref{alg:psiapf:second_weight_update:1}--\ref{alg:psiapf:second_weight_update:2} from Step~($t-1$) results exactly in the \gls{PSIAPF} as presented in \citet{amj26:GJL17, smc:methodology:HBDD20}.

Thirdly, Lines~\ref{alg:psiapf:resample:1}--\ref{alg:psiapf:resample:4} perform adaptive resampling \citet{kong1994sequential,liu1995blind} based on the weights $V^{1:\nParticles}$. That is, whenever the \emph{\gls{ESS}} drops below $\kappa N$, for some user-defined threshold $0 < \kappa \leq 1$, $\resample{V^{1:\nParticles}}$ samples the ancestor indices $A_{t-1}^{1:\nParticles}$ via some suitable resampling scheme based on a set of weights $V^{1:\nParticles}$; numerous resampling schemes have been proposed over the years, in our numerical experiments we use residual--multinomial resampling \citep{liu1998sequential}.

Finally, there exists an optimal sequence $\psi$ which yields zero-variance estimates of final-time normalising constants $Z_T^N$ \citep[Proposition 2]{amj26:GJL17}, but this sequence is generally not computable or usable in practice. We are constrained in the models to which this approach can be applied and the twisting functions that can be employed by the fact that we must be able to sample from $f^\psi_t$ for each $t$ and also to evaluate the corresponding $g^\psi_t$; one surprisingly common setting in which this is possible is that in which the transition densities are Gaussian and the twisting functions employed are exponentials of a quadratic form.

\subsection{(Offline) Controlled SMC} 
\glsreset{CSMC}

The \emph{\gls{CSMC}} algorithm---also known as \emph{iterated \gls{APF}}--introduced by \cite{amj26:GJL17, smc:methodology:HBDD20}, excels in offline contexts by leveraging a complete data set $y_{1:T}$ to provide accurate approximations of smoothing distributions of the form $p(x_{1:T} | y_{1:T})$ and the marginal likelihood $p(y_{1:T})$ for a class of models. 

A key insight underlying \gls{CSMC} is that the problem of incorporating the influence of future observations can be cast as a function approximation problem at each time point. In particular, \citet{smc:methodology:HBDD20} cast this problem as one of optimal control in which one seeks to learn the policy. As summarised in Algorithm~\ref{alg:csmc}, the idea is to repeatedly apply the \gls{PSIAPF} (from time~$1$ to time~$T$), learning a better sequence of twisting functions during each iteration to use it in the next. 

\begin{algorithm}[htbp]
\caption{\acrfull{CSMC}}
\label{alg:csmc}
\begin{algorithmic}[1]
  \Require Initial particle system $\particleSystem{0}$ as in \eqref{eq:initial_particle_system}.
  \State Set $\twistingFunction{1} = \dotsc = \twistingFunction{T+1} \equiv 1$.
  \For {$k = 1, \dotsc, K$}
    \For {$t = 1, \dotsc, T$}
      \State sample $\particleSystem{t} \leftarrow \psiapf{t}$, \Comment{Algorithm~\ref{alg:psiapf}}
    \EndFor
    \For {$t = T, \dotsc, 1$}
      \State set
      $\twistingFunction{t} \leftarrow \learnpsi{t}$. \Comment{Algorithm~\ref{alg:learn_psi}}
    \EndFor
  \EndFor
  \Ensure Approximations $\smash{p(y_{1:T}) \approx Z_T^\nParticles}$ and $p(x_{1:T}|y_{1:T}) \approx\smash{\sum_{n=1}^\nParticles W_T^n} \delta_{X_{1:T}^{(n)}}$ (based on $\particleSystem{1:T}$ available at the end of the algorithm), where $\smash{X_{1:T}^{(n)}}$ denotes the $n$th particle lineage at time $T$, recursively defined as $\smash{X_{1:t}^{(n)} \coloneqq (X_{1:(t-1)}^{(A_{t-1}^n)}, X_t^n)}$, with initial condition $X_1^{(n)} \coloneqq X_t^n$. 
\end{algorithmic}
\end{algorithm}

\begin{algorithm}[htbp]
\caption{$\learnpsi{t}$}
\label{alg:learn_psi}
\begin{algorithmic}[1]
 \Require Time index $t \in \naturals$.
 \Require Twisting function $\twistingFunction{t+1}$.
 \Require Particle system $\particleSystem{t}$ as in \eqref{eq:particle_system}.
    \State\label{alg:learn_psi:psit}Set $\twistingFunction{t}^n \leftarrow g_t(y_t | X_t^n) \transitionIntegral{t+1}{X_t^n}$,
        for $n \in \dcount{N}$.
    \State\label{alg:learn_psi:opti}Choose $\twistingFunction{t}$ on the basis of
        $X_t^{1:\nParticles}$ and $\twistingFunction{t}^{1:\nParticles}$.
 \Ensure $\psi_t$.
  \end{algorithmic}
\end{algorithm}
The precise implementation of Line~\ref{alg:learn_psi:opti} in Algorithm~\ref{alg:learn_psi} varies between \citet{amj26:GJL17} and \citet{smc:methodology:HBDD20}. Both methodologies converge on solving an optimization problem which is designed to refine the current policy towards an optimal one, utilizing information obtained from current sample values. Our specific approach is detailed in Section~\ref{subsec:setup} below.

Empirical results have demonstrated that \gls{CSMC} performs significantly better than the standard \gls{BPF} in challenging scenarios, and can significantly outperform the fully-adapted \gls{APF}, often regarded as a gold standard in the online setting, at least in some simple models for which it can be implemented. 

\section{Online Rolling Controlled SMC}\label{sec:cops}
\glsreset{ORCSMC}

Unfortunately, the application of \gls{CSMC} in online settings is limited because it requires processing of the whole data sequence. Although warm-starting would be possible and one could simply re-run the entire algorithm as observations become available, the cost of processing each new observation would grow with the length of the data sequence, making the computational complexity of such an approach prohibitive.

In this section, we present an online \gls{CSMC} scheme---termed \emph{\gls{ORCSMC}} and summarised in Algorithm \ref{alg:orcsmc}---that transfers the high accuracy of \gls{CSMC} to real-time inference through a rolling-window, dual-filter design scenario. Specifically, the underlying strategy is to consider two concurrent \glspl{PSIAPF} at time $t$: 
\begin{compactenum}
    \item The \emph{learning filter} (Lines~\ref{alg:orcsmc:learning_filter:1}--\ref{alg:orcsmc:learning_filter:2}) (whose particle systems are indicated by a `$\sim$'-accent) repeatedly runs over the rolling window $\dinterval{t_0}{t}$, where $t_0 \coloneqq \max\{1, t-L+1\}$, to obtain approximations, $\twistingFunction{t_0:t}$, of optimal twisting functions. These updates are warm-started by initialising $\twistingFunction{t_0:(t-1)}$ to the twisting functions obtained based on the learning filter at time $t-1$; $\twistingFunction{t}$ is initialised to the unit function.
    \item The \emph{estimation filter} (Line~\ref{alg:orcsmc:estimation_filter:1}) runs over the time steps $s = t_0, \dotsc, t$ using the latest values of the twisting functions $\twistingFunction{t_0:t}$ obtained in Line~\ref{alg:orcsmc:learn_psi}. Note that it thus overwrites the values of $\particleSystem{t_0:(t-1)}$ it had generated at time $t-1$. The filtering estimates output by the algorithm at time $t$ are then computed using the quantities $\smash{X_t^{1:\nParticles}}$, $\smash{W_t^{1:\nParticles}}$, $\smash{A_{t-1}^{1:\nParticles}}$ and $Z_t^\nParticles$ which are assumed to come from the (most recent values of the) particle systems $\particleSystem{1:t}$ available at time $t$. Specifically, $Z_t^\nParticles$ obtained from $\particleSystem{t}$ in Line~\ref{alg:orcsmc:output} is an unbiased estimate of $p(y_{1:t})$. If the user requires filtering estimates only at \emph{some}  times $\mathcal{T} \subseteq \dcount{T}$, then the above-mentioned `overwriting' can be skipped at other times. More precisely, the sampling operation in Line~\ref{alg:orcsmc:estimation_filter:1} then (a) only needs to be fully performed if $t \in \mathcal{T} \cup \{T\}$; (b) can be skipped entirely if $t < L$ and $t \notin \mathcal{T}$; (c) only needs to be carried out for $s = t_0 = t - L + 1$, at all other times $t$. 
\end{compactenum}

\begin{algorithm}
\caption{\acrfull{ORCSMC}} \label{alg:orcsmc}
\begin{algorithmic}[1]
 \Require Time horizon $T \in \naturals \cup \{\infty\}$.
 \Require Initial particle systems $\particleSystem{0} = \particleSystemAlt{0}$ whose components are defined in \eqref{eq:initial_particle_system}.
    \State Set $\twistingFunction{1} = \dotsc = \twistingFunction{T+1} \equiv 1$.
    \For {$t = 1, \dotsc, T$} 
      \State set $t_0 \coloneqq \max\{1, t - L + 1\}$,
      \If {$t - 1 > L$} 
        \State\label{alg:orcsmc:discard}discard $\twistingFunction{t-L-1}$, $\particleSystemAlt{t-L-1}$, $\particleSystem{t-L-1}$, \Comment{optional}
      \EndIf
      \State\label{alg:orcsmc:learning_filter:1}sample $\particleSystemAlt{t} \leftarrow \psiapfAlt{t}$; \Comment{Algorithm~\ref{alg:psiapf}}
      \For {$k = 1, \dotsc, K$}
          \For {$s = t, t - 1, \dotsc, t_0$} 
            \State\label{alg:orcsmc:learn_psi}set $\twistingFunction{s} \leftarrow \learnpsiAlt{s}$, \Comment{Algorithm~\ref{alg:learn_psi}}
          \EndFor
         \For{$s = t_0, \dotsc, t$}
          \State\label{alg:orcsmc:learning_filter:2}sample
          $\particleSystemAlt{s} \leftarrow \psiapfAlt{s}$; \Comment{Algorithm~\ref{alg:psiapf}}
          \EndFor
        \EndFor
        \For {$s = t_0, \dotsc, t$}
        \State\label{alg:orcsmc:estimation_filter:1}sample $\smash{\particleSystem{s} \leftarrow \psiapf{s}}$; \Comment{Algorithm~\ref{alg:psiapf}}
        \EndFor
        \State\label{alg:orcsmc:output} \textbf{Output:} Approximations 
          $\smash{p(y_{1:t}) \approx Z_t^\nParticles}$ and $p(x_{1:t}|y_{1:t}) \approx\smash{\sum_{n=1}^\nParticles W_t^n} \delta_{X_{1:t}^{(n)}}$ (based on $\particleSystem{1:t}$). Here, $\smash{X_{1:t}^{(n)} \coloneqq (X_{1:(t-1)}^{(A_{t-1}^n)}, X_t^n)}$ is recursively defined with initial condition $\smash{X_1^{(n)} \coloneqq X_t^n}$.
    \EndFor
    \end{algorithmic}
\end{algorithm}
Line~\ref{alg:orcsmc:discard} of Algorithm~\ref{alg:orcsmc} is marked as `optional' because it is only needed to ensure that the memory cost is bounded in $t$. Of course, this implicitly assumes that we are not interested in outputting approximations of (marginals of) $p(x_{1:(t-L-1)}|y_{1:t})$ in Line~\ref{alg:orcsmc:output}.

This framework is roughly as flexible as the (offline) \gls{CSMC} algorithm which it extends; see Section~\ref{subsec:setup} for details of the implementation employed in our empirical study. 

\section{Experiments}\label{sec:experiments}
In this section, we illustrate that a number of models used to illustrate the behaviour of offline algorithms by \citet{amj26:GJL17,smc:methodology:HBDD20} also admit efficient online inference using the controlled approaches developed in this paper.

\subsection{Setup}
\label{subsec:setup}
Throughout all experiments, we assume homogeneous linear-Gaussian dynamics with $\mu(x_1) \coloneqq \dnorm(x_1; m, \varSigma)$ and  $f_t(x_t|x_{t-1}) \coloneqq \dnorm(x_t; A x_{t-1}, B)$, for $m \in \reals[d]$, $A, B, \varSigma \in \reals[{d \times d}]$ with $B$ and $\varSigma$ positive definite.

To iteratively optimize the twisting functions $\twistingFunction{t}$ within both \gls{CSMC} and \gls{ORCSMC}, we use the \emph{\gls{ADP}} approach from \citet{smc:methodology:HBDD20} to implement Line~\ref{alg:learn_psi:opti} of Algorithm~\ref{alg:learn_psi}. This entails minimising a squared error on the logarithmic scale based on the terms $\twistingFunction{t}^n$ from Line~\ref{alg:learn_psi:psit} of Algorithm~\ref{alg:learn_psi}. 
For this purpose, our experiments adopt a quadratic function class for optimization: the family of functions of the form $\psi_t(x_t) = \exp(x^T_t \mathsf{A}_t
x_t + \mathsf{b}_t^T x_t + \mathsf{c}_t)$ for $\mathsf{A}_t \in \mathbb{S}_d = \{\mathsf{A} \in \reals[d \times d] : \mathsf{A} = \mathsf{A}^T\}$, $\mathsf{b}_t \in \reals[d]$ and $\mathsf{c}_t \in \reals$ and $x_t$ is a multivariate state vector. By treating the negative logarithm of the \gls{ADP} target values as the dependent variable, we transform the optimization into a linear least squares problem. The coefficients $\mathsf{A}_t$ and $\mathsf{b}_t$ derived from this regression then directly forms the parameters of a Gaussian density, which serves as our optimized policy, defining the learned $\twistingFunction{t}$. To ensure that the number of parameters of the twisting functions grows only linearly with $d$, we restrict the matrices $\mathsf{A}_t$ to be diagonal.

Throughout all our experiments, both \gls{CSMC} and \gls{ORCSMC} use $K = 5$ iterations for learning the twisting functions. We focus in particular on the  estimates $Z_T^N$ of the final-time normalising constant $Z_T = p(y_{1:T})$ as these are well known to be a good indicators of the approximation quality of \glspl{PF}. All results are based on \num{100} independent replicates.

\subsection{Linear-Gaussian Model}
To illustrate the behavior of \gls{ORCSMC}, we begin with a simple linear-Gaussian toy model in which $g_t(\ccdot|x_t)= \dnorm(\ccdot ; C x_t, D)$, where $m = \mathbf{0}_d$ is the zero vector in $\reals[d]$, and $\varSigma = B = C = D = I_d$, where $I_d$ is the $(d \times d)$-identity matrix. This model admits exact inference via the Kalman filter and hence provides a convenient testbed. 

Throughout, we fix $T = 100$ and employ $N = \num{1000}$ particles for inference within \gls{ORCSMC}. We also consider state dimensions $d \in \{2, 4, 8, 16, 32, 64\}$ and lags $L \in \{2, 4, 8, 16\}$. To balance the computational cost, where comparisons are conducted, we use $N = \num{320000}$ particles for the \gls{BPF} and $N = \num{14000}$ for \gls{CSMC}. Since $Z_T = p(y_{1:T})$ is analytically tractable in this model, we consider the more meaningful \emph{relative} estimates of the normalising constant, $Z_T^N / Z_T$, throughout.

We consider two different linear Gaussian models to explore the impact of restricting the class of twisting functions to a diagonal class, where $\alpha \coloneqq 0.415$:
\begin{enumerate}
    \item in the \emph{diagonal case,} we set $A = \alpha I_d$; 
    \item in the \emph{non-diagonal case,} we follow \citet{amj26:GJL17}, in setting $[A]_{ij} = \alpha^{|i-j|+1}$, for $(i,j) \in \dcount{d}^2$.
\end{enumerate}

\subsubsection{Diagonal Case}

We first consider the case that the diagonal structure of the transition matrix $A = 0.415 I_d$ is compatible with the form of the twisting function $\psi_t$ (recall that we restrict the parameter $\mathsf{A}_t$ of $\twistingFunction{t}$ to be a diagonal matrix).

\begin{figure}
  \centering
  \includegraphics[width=0.75\linewidth]{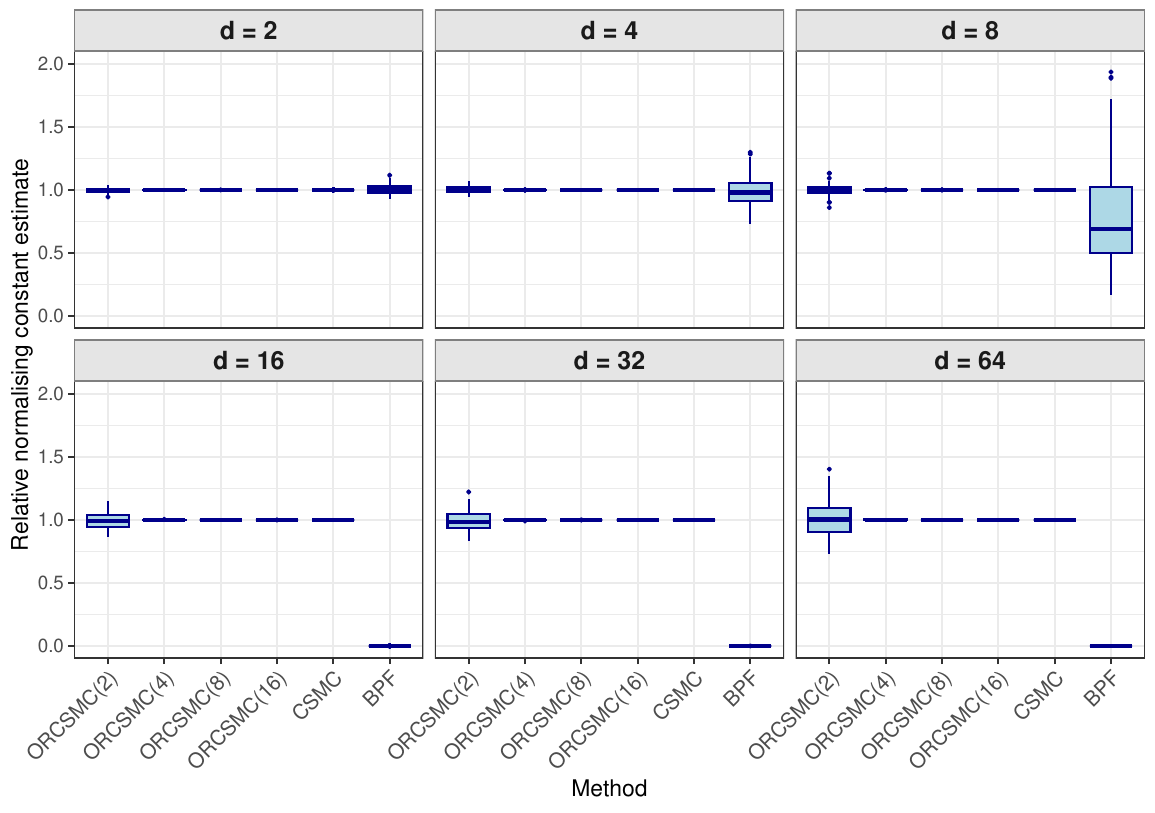}  
  \caption{Relative normalising constant estimates for the diagonal Gaussian model obtained via \gls{ORCSMC} (where parenthetical numbers indicate lag), \gls{CSMC} and \gls{BPF}.}  
  \label{fig:nd_diag}
\end{figure}

Figure~\ref{fig:nd_diag} compares the normalising constant estimates obtained by \gls{ORCSMC}, the \gls{BPF} and the (offline) \gls{CSMC} algorithm.

\subsubsection{Non-diagonal Case}

We now consider the case that the transition matrix $A$ has the above-mentioned non-diagonal structure previously used in \citet{amj26:GJL17}, which is less favourable to the diagonal twisting functions which we employ.

\begin{figure*}[htbp]
  \centering
  \includegraphics[ height=8cm]{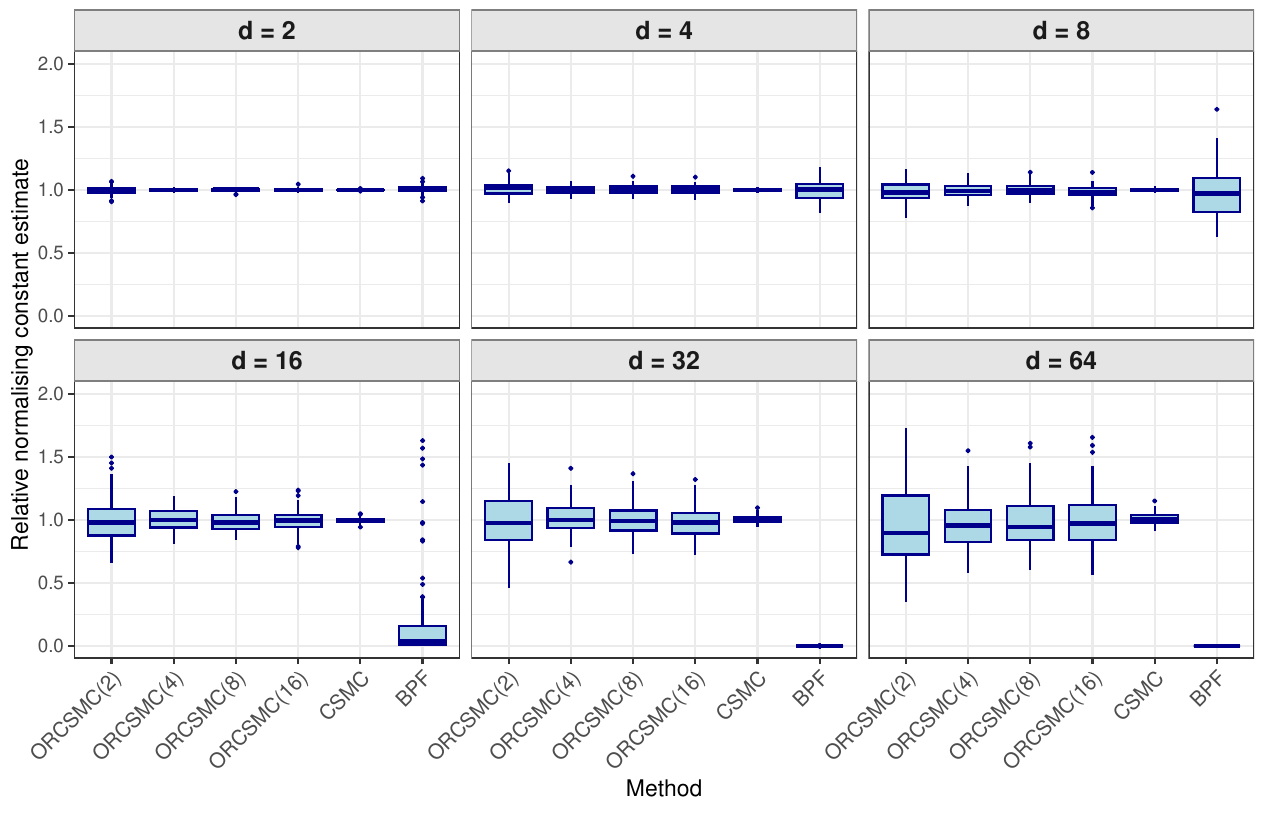} 
  \caption{Relative normalising constant estimates for the non-diagonal Gaussian model obtained via \gls{ORCSMC} (where parenthetical numbers indicate lag), \gls{CSMC} and \gls{BPF}.}
  \label{fig:nc}
\end{figure*}

Figure~\ref{fig:nc} illustrates that \gls{ORCSMC} outperforms the \gls{BPF} across nearly all dimensions in terms of both accuracy and stability of the normalising-constant estimates. 

While \gls{CSMC} exhibits some improvement over \gls{ORCSMC}, this advantage is marginal if a sufficient lag length is used by the latter, and is expected as offline methods inherently offer certain benefits over their online counterparts.

\begin{figure}[htbp]
    \centering
    \includegraphics[width=0.75\textwidth]{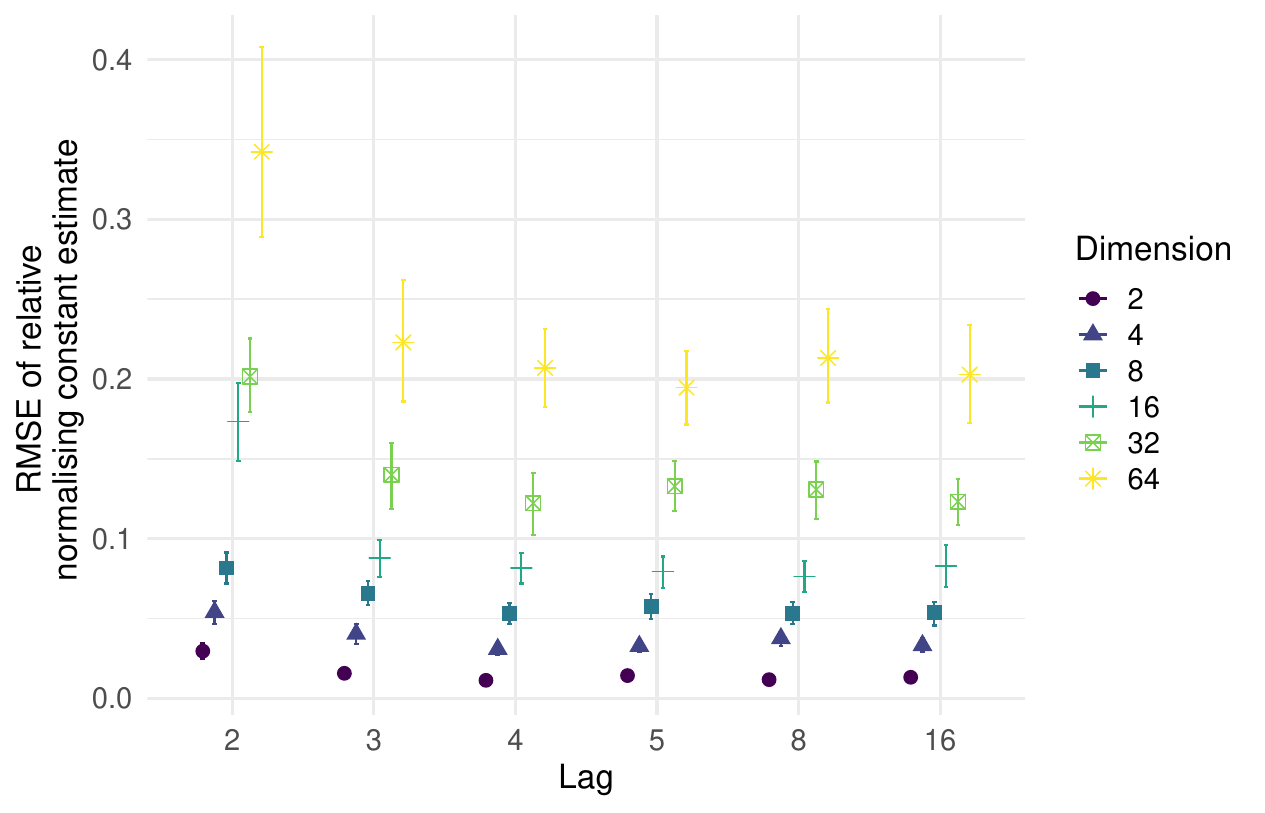}
    \caption{\Gls{RMSE}, computed across the \num{100} independent replicates, of the normalising constant estimates for the non-diagonal Gaussian model in different dimensions.}    
    \label{fig:rmse}
\end{figure}

Figure~\ref{fig:rmse} indicates that the approximation error of \gls{ORCSMC} decays with the lag; and this finding appears to hold across all considered dimensions.

\begin{figure}[htbp]
  \centering
  \includegraphics[width=0.8\textwidth]{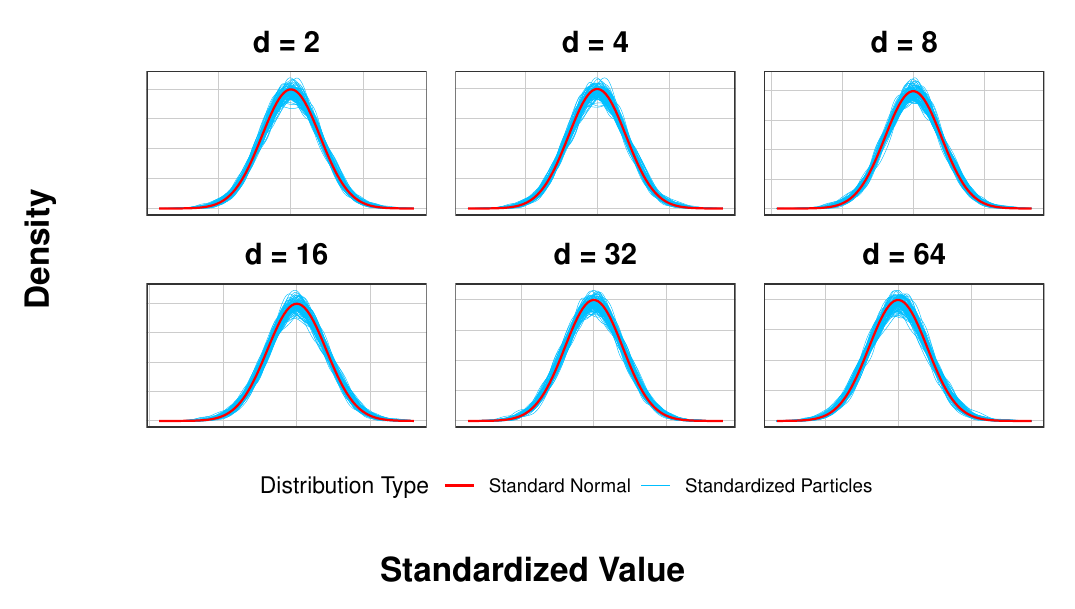}  
  \caption{The distribution comparison between the standardized first coordinate marginals of the empirical distribution and the optimal smoother in the non-diagonal Gaussian model at each time $t$---obtained from a single run of \gls{ORCSMC} for each value of $d$.}
  \label{fig:smoo}
\end{figure}

Figure~\ref{fig:smoo} investigates how accurately \gls{ORCSMC} approximates the marginal smoothing distribution $p(x_t|y_{1:T})$ at time $t$. Specifically, at each time $t$, the true mean and standard deviation of the first coordinate of the state variable, computed using a Kalman filter--smoother, are used to standardize the empirical distribution obtained from the particles generated by \gls{ORCSMC}. The resulting standardized empirical distribution is then compared with the standard normal distribution. This plot reveals a high degree of concordance between these two distributions, indicating that the algorithm is capable of providing accurate smoothing results.

\begin{figure}[htbp]
  \centering
  \includegraphics[width=0.75\textwidth]{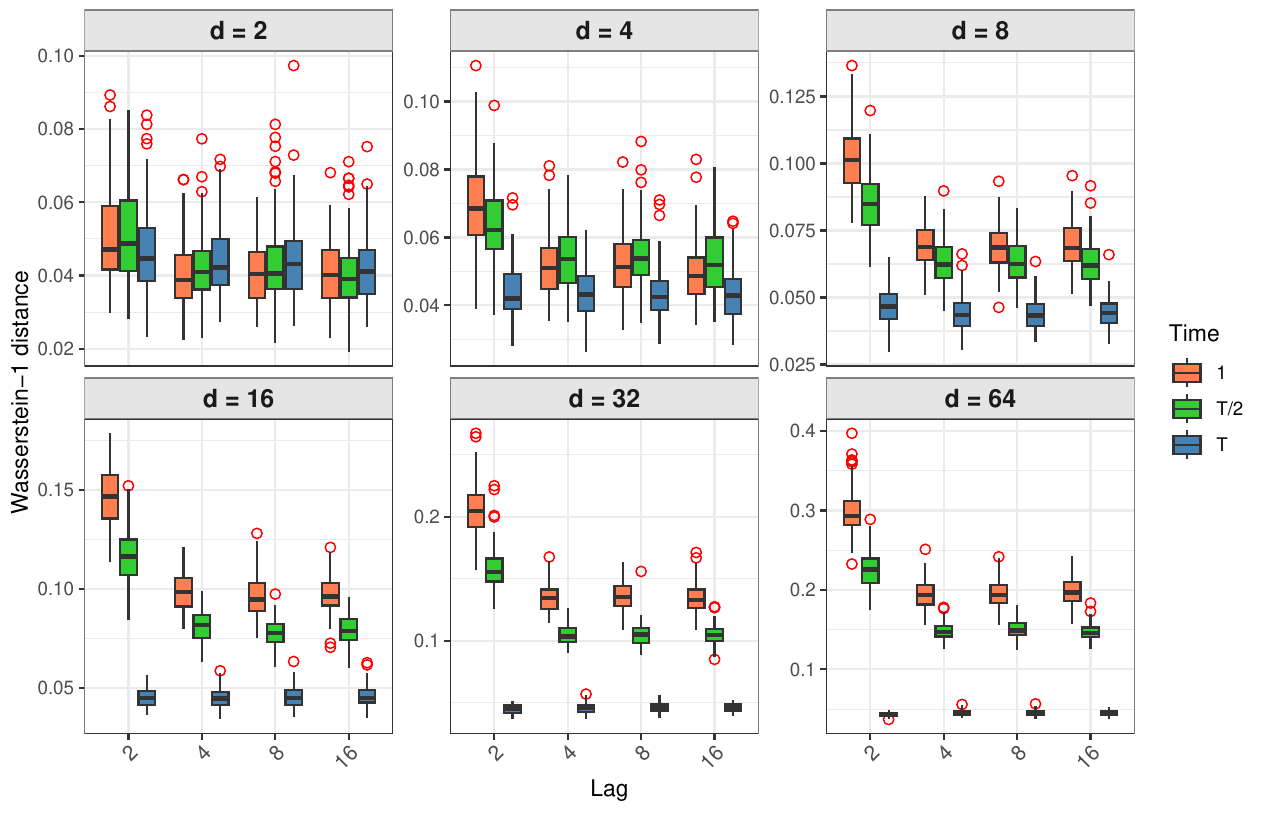}  
  \caption{The Wasserstein-1 distance between empirical and exact smoothing distributions in the non-diagonal Gaussian model. }
  \label{fig:l1}
\end{figure}

Figure~\ref{fig:l1} compares the empirical marginal \gls{CDF} $\hat{F}_{t,j}$ to the true Gaussian \gls{CDF} $F_{t, j}$ (of $p(x_t|y_{1:T})$) at times $t \in \{1, T/2, T\}$. For each state coordinate $j \in \dcount{d}$, this comparison is made using the $L_1$ distance between their cumulative distribution functions, which coincides with the Wasserstein-1 distance between their distributions via the Kantorovich--Rubinstein duality:
\begin{align}
    L_{1}(t, j)
      = 
    \int_{-\infty}^{\infty}
    \bigl\lvert \hat{F}_{t,j}(x) - F_{t,j}(x)\bigr\rvert \intDiff x.
\end{align}
Thus, $L_{1}(t, j)$ represents the Wasserstein-1 distance between the empirical distribution function of the $j$th coordinate of the particle approximation of $p(x_t|y_{1:T})$ and its true (Gaussian) distribution at time $t$. The results show stability over time; the average marginal $L_1$ error, calculated as the mean of $L_{1}(t, j)$ across all state coordinates $j \in \dcount{d}$, remains stable for all $t$, which indicates accurate marginal smoothing performance across coordinates.

\subsection{Stochastic Volatility Model}
We consider a simple univariate stochastic volatility model as in \cite{kim1998stochastic}, $\mu(\ccdot) = \dnorm(\ccdot; 0, \sigma^2 / (1 - \alpha^2))$, $f_t(\ccdot|x_t) = \dnorm(\cdot; \alpha x_t, \sigma^2)$ and $g_t(\ccdot|x_t) = \dnorm(\cdot; 0,\beta^2 \exp(x_t))$
for $\alpha \in (0, 1)$, $\beta > 0$ and $\sigma^2 > 0$. The parameters were estimated from weekday close exchange rates data from 1/10/81 to 28/6/85 ($T = 945$ observations) by taking the approximate posterior mode (under a weakly informative prior) from \citet[Figure~5]{amj26:GJL17} giving $\alpha = 0.986, \sigma = 0.13$ and $\beta = 0.69$; these values are also close to the MLE provided in the same paper. We run the \gls{ORCSMC} with $N = \num{200}$ particles.

The experiment shows that even in this non-Gaussian, strongly heteroskedastic setting, \gls{ORCSMC} is capable of producing stable marginal‐likelihood estimates. As the lag $L$ increases, the variance of (the logarithm of) $Z_T^\nParticles$ falls substantially, demonstrating its robustness for stochastic volatility models.

\begin{figure}
    \centering
    \includegraphics[width=0.6\textwidth]{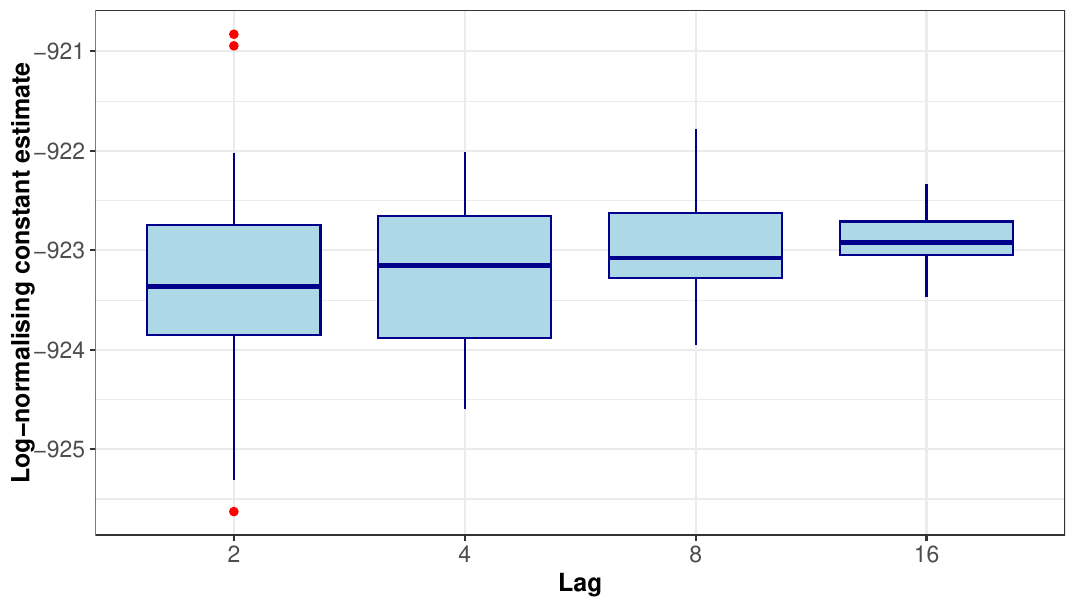}
    \caption{The normalising constant estimates for the stochastic volatility model.}
    \label{fig:svm}
\end{figure}

\subsection{Neuroscience Model} 

Finally, we evaluate \gls{ORCSMC} using a neuroscience-inspired model, following the setup in \cite{smc:methodology:HBDD20}. This defines a time-homogeneous state-space model on $\reals{}$. At each time step $t$, the observation $y_t \in [0,M]$ represents the number of activated neurons out of $M = 50$ repeated experiments. The observation likelihood for the univariate case is given by $g_t(y_t|x_t) = \mathrm{Bin}(y_t; M, \kappa(x_t))$, where $\kappa(\cdot)$ is the logistic link function, defined as $\kappa(z) \coloneqq 1 / (1 + \exp(-z))$. The dynamics are specified as $\mu(\ccdot) \coloneqq \dnorm(\ccdot; 0, 1)$ and $f_t(\ccdot|x_{t-1}) \coloneqq \dnorm(\ccdot; \alpha x_{t-1}, \sigma^2)$, for $\alpha = 0.99$ and $\sigma^2 = 0.11$.

To allow us to also investigate the behaviour in higher dimensions, we extend this model to a multivariate setting in which both the states $X_t \coloneqq X_{t, 1:d}$ and observations $Y_t \coloneqq Y_{t,1:d}$ are $d$-dimensional. To that end, we set $m = \mathbf{0}_d, \varSigma = I_d, A = \alpha I_d, B = \sigma^2I_d$. The observation densities are $g_t(y_t|x_t) = \prod_{j=1}^d \mathrm{Bin}(y_{t,j}; M, \kappa(x_{t,j}))$.

\begin{figure}[htbp]
\begin{subfigure}[t]{0.485\textwidth}
    \centering
    \includegraphics[width=\textwidth]{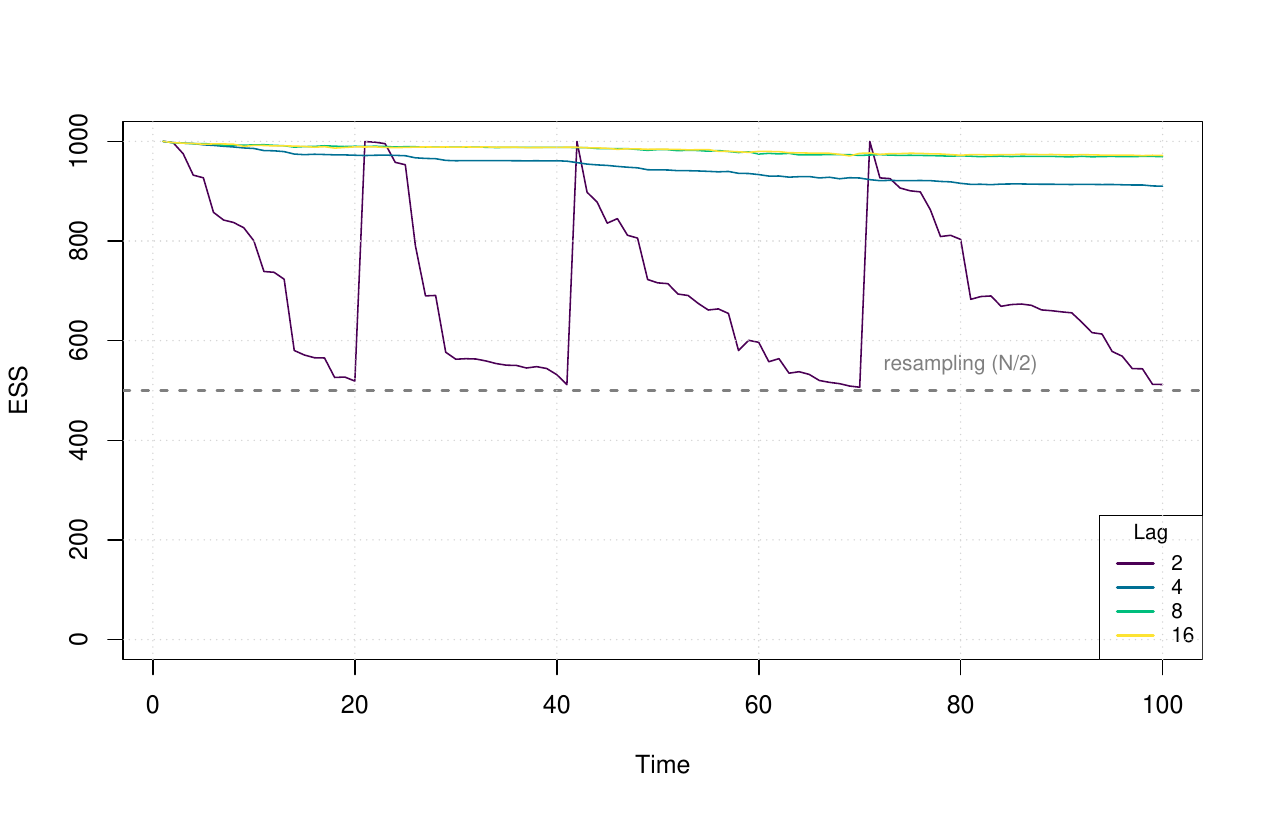}
    \caption{Evolution of \gls{ESS} over time.
    }
    \label{fig:ess}
\end{subfigure}\hfill%
\begin{subfigure}[t]{0.485\textwidth}
    \centering
    \includegraphics[width=\textwidth]{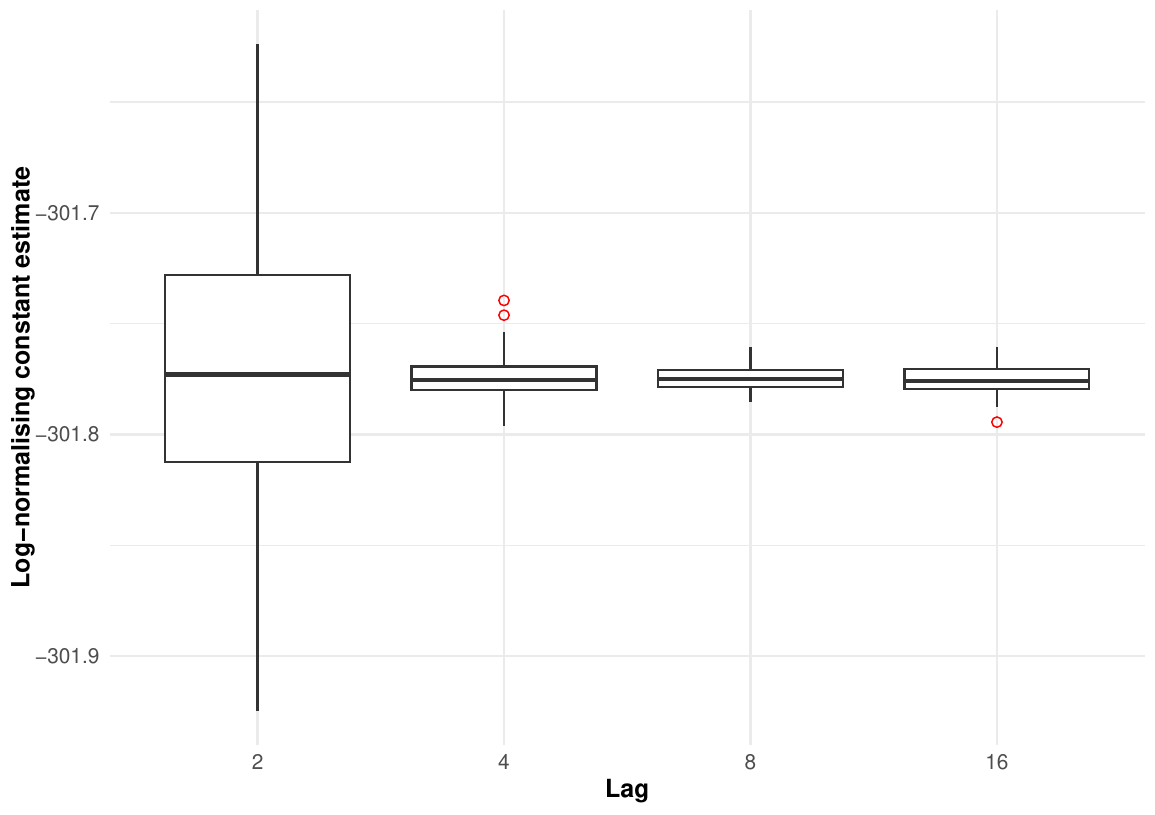}
    \caption{Log-normalising constant estimates. 
    }
    \label{fig:bin}
\end{subfigure}
\caption{Results for univariate binomial model for various lags.}\label{fig:unibinom}
\end{figure}

Figures~\ref{fig:unibinom} and \ref{fig:bin} illustrate the performance of \gls{ORCSMC} in this model in terms of the evolution of the \gls{ESS} over time and the stability of log-normalising constant estimates.

\begin{figure}[htbp]
    \centering
    \includegraphics[width=0.8\textwidth]{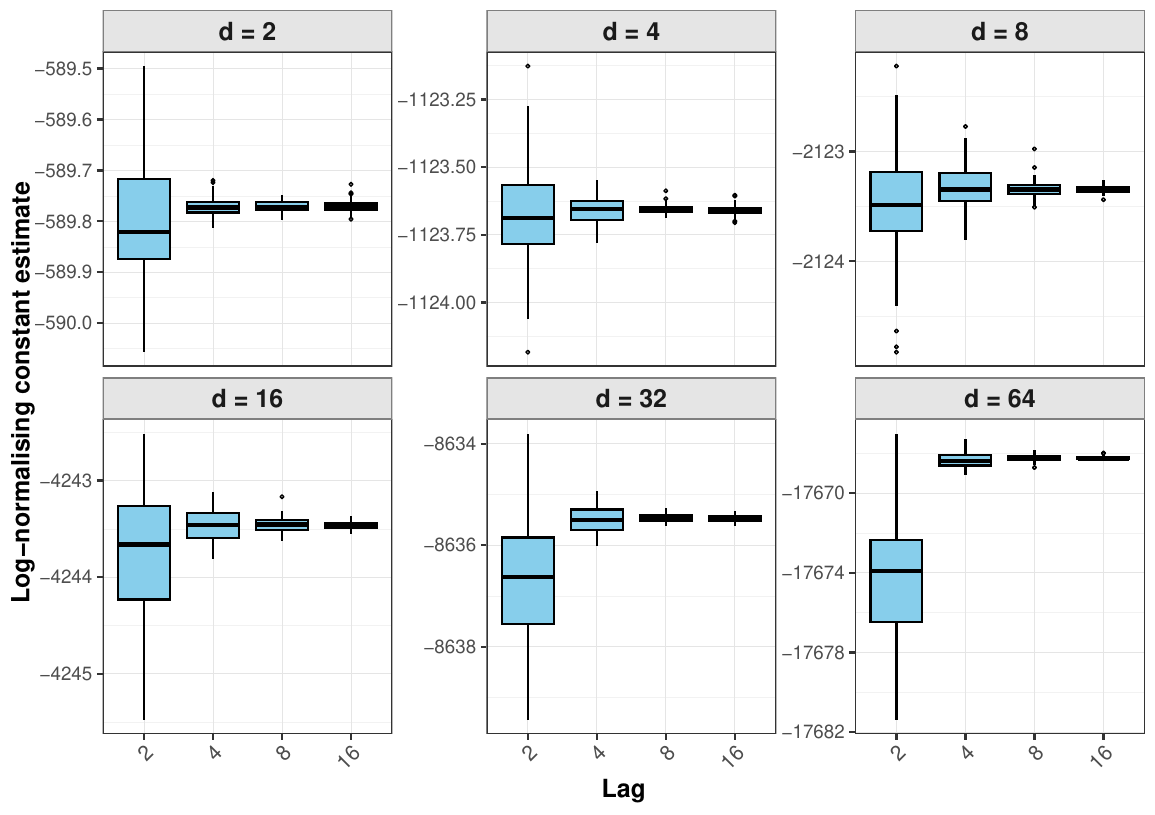}
    \caption{Log-normalising constant estimates obtained from \gls{ORCSMC} (with different lags) for the multivariate binomial model in different dimensions. 
    }     
    \label{fig:bin}
\end{figure}

\section{Discussion}\label{sec:conclusion}
We have presented an approach for performing online inference in a class of state-space models which exploits adaptive guiding methods. It extents the applicability of \gls{CSMC} methods to settings in which one wishes to approximate filtering or smoothing distributions, or associated normalising constants, online as observations become available.

Proof-of-concept numerical results illustrate the performance of the method on some linear Gaussian toy models, a stochastic volatility model and a model inspired by a problem in neuroscience. Within the family of models with Gaussian transitions for which the particular numerical method used herein might find use lie many models in which the underlying dynamics are obtained from the time discretisation of an underlying stochastic differential equation (see \cite{guarniero2017} for the case of simple Euler--Maruyama discretisations or \cite{huang2025} for approaches based on splitting schemes and diffusion bridges) and the approach developed here should allow good \emph{online} inference for those models.

As with all \gls{CSMC} methods in the literature, the breadth of applicability of the method depends upon the class of models which can be handled. While the randomization approach of \cite{bon2022monte} provides one avenue to extend the applicability of methods, and exploiting broader conjugacy properties than those of Gaussian families (such as those described in \cite{smc:hmm:Vid99}) would be feasible, there is scope for further development in this area.

\if\version2
\section*{Acknowledgements}
\actualacknowledgements
\fi
 
\bibliographystyle{jabes}

\bibliography{quasi-online}
\end{document}

%% file: abbreviations.tex
\newacronym{PMCMC}{PMCMC}{particle Markov chain Monte Carlo}%
\newacronym{APF}{APF}{auxiliary particle filter}%
\newacronym{PSIAPF}{$\psi$-APF}{$\psi$-auxiliary particle filter}%
\newacronym{FAAPF}{FA-APF}{fully-adapted auxiliary particle filter}%
\newacronym{PF}{PF}{particle filter}%
\newacronym{HMM}{HMM}{hidden Markov model}%
\newacronym{PDF}{PDF}{probability density function}%
\newacronym{IID}{IID}{independent and identically distributed}%
\newacronym{MCMC}{MCMC}{Markov chain Monte Carlo}%
\newacronym{MH}{MH}{Metropolis--Hastings}%
\newacronym{ESS}{ESS}{effective sample size}%
\newacronym{CDF}{CDF}{cumulative distribution function}%
\newacronym{SMC}{SMC}{sequential Monte Carlo}%
\newacronym{ORCSMC}{ORCSMC}{online rolling controlled SMC}%
\newacronym{CSMC}{CSMC}{controlled sequential Monte Carlo}%
\newacronym{EPSRC}{EPSRC}{Engineering and Physical Sciences Research Council}%
\newacronym{CLT}{CLT}{central limit theorem}%
\newacronym{BPF}{BPF}{bootstrap particle filter}%
\newacronym{CI}{CI}{confidence interval}%
\newacronym[user1={importance-sampling}]{IS}{IS}{importance sampling}%
\newacronym{MSE}{MSE}{mean-square error}%
\newacronym{KL}{KL}{Kullback--Leibler}%
\newacronym{RMSE}{RMSE}{root mean-square error}%
\newacronym{MLE}{MLE}{maximum-likelihood estimate}%
\newacronym{ML}{ML}{maximum-likelihood}%
\newacronym{IAPF}{iAPF}{iterated auxiliary particle filter}%
\newacronym{SV}{SV}{stochastic volatility}%
\newacronym{ADP}{ADP}{approximate dynamic programming}